\documentclass[12pt]{article}
\usepackage{graphicx,amssymb,amsfonts,amsmath,epsfig}

\renewcommand{\theequation}{\arabic{section}.\arabic{equation}}

\newcommand{\gapprox}{%
\mathrel{%
\setbox0=\hbox{$>$}\raise0.6ex\copy0\kern-\wd0\lower0.65ex\hbox{$\sim$}}}
\newcommand{\be}{\begin{equation}}
\newcommand{\ee}{\end{equation}}
\newcommand{\bea}{\begin{eqnarray}}
\newcommand{\eea}{\end{eqnarray}}
\newcommand{\cx}{\overset{\circ}{x}_2}
\def\CN{$\mathcal{N}$}

\def\t{\theta}
\newcommand{\bref}[1]{(\ref{#1})}

\def\avall{\vspace{0.5cm}}

\begin{document}

\begin{titlepage}

\setcounter{page}{0}
\setcounter{table}{0}
\setcounter{figure}{0}
\begin{flushright}
Imperial/TP/2-03 /21\\
\end{flushright}

\vspace{5mm}
\begin{center}
{\Large {\bf On the consistency of the \CN=1 SYM spectra
from wrapped five-branes}} \vspace{10mm}

{\large
Llu{\'\i}s Ametller$^a$, Josep M. Pons$^{b,c}$ and Pere Talavera$^{a}$} \\
\vspace{5mm}
$^a${\em
Departament de F{\'\i}sica i Enginyeria Nuclear,
Universitat Polit\`ecnica de Catalunya,\\
Jordi Girona 1--3, E-08034 Barcelona, Spain.}\\[.1cm]
$^b${\em
Departament d'Estructura i Constituents de la Mat\`eria,
Universitat de Barcelona,\\
Diagonal 647, E-08028 Barcelona, Spain.}\\[.1cm]
$^c${\em Theoretical Physics Group,
Blackett Laboratory, Imperial College London, SW7 2BZ, U.K.}\\[.1cm]
\vspace{5mm}
\end{center}

\vspace{20mm} \centerline{{\bf{Abstract}}}
\avall
We discuss the
existence of glueball states for \CN=1 SYM within the
Maldacena-N\'u\~nez model. We find that for this
model the existence of an area law in the Wilson loop operator
does not imply the existence of a discrete glueball spectrum. We suggest
that implementing the model with an upper hard cut-off
can amend the lack of spectrum. As a result the model can be \emph{only}
interpreted in the infra-red region.
A direct comparison with the lattice data allows us to fix the
scale up to where the model is sensible to describe low-energy
observables.
Nevertheless, taking for granted the lattice results,
the resulting spectrum does not
follow the general trends found in other supergravity backgrounds.
We further discuss the decoupling of the non-singlet
Kaluza--Klein states by analysing the associated
supergravity equation of motion. The inclusion of non-commutative
effects is also analysed and we find that leads to
an enhancement on the value of the masses.
\vspace{5mm}
\vfill{
 \hrule width 5.cm
\vskip 2.mm
{\small
\noindent E-mail: lluis.ametller@upc.es, pons@ecm.ub.es, pere.talavera@upc.es
}}

\end{titlepage}

\newpage

\tableofcontents       %
\vskip 1cm             %

\setcounter{equation}{0}
\section{Motivation}
\label{Mot}

The success of the gauge/string correspondence is best understood
in simple backgrounds that are typically both supersymmetric and
conformally invariant \cite{conjecture1}. The precise gauge/string
relation determines the dimension of the conformal field operators
in terms of the masses of particles in the string theory
\cite{conjecture2,conjecture3}.  Aiming to a dual description of
non-perturbative QCD, it is reasonable to extend the correspondence
to models in which both symmetries are partially reduced or
eventually eliminated, thereby obtaining properties of
non-supersymmetric gauge theories in the large-$N$ limit. According
to \cite{sfetsos} the supergravity duals of
non-supersymmetric gauge theories are non-supersymmetric
background solutions in type--IIB theory. {F}or instance, one can
easily extend the duality to discuss the large-$N$ behaviour of
QCD in 3 and 4 dimensions \cite{witten} by compactifying one of
the (Euclidean) space-time directions of AdS on a circle, in which
case supersymmetry is broken by the choice of antiperiodic
boundary conditions of the circle for the fermions.

With the help of this and similar models some QCD-like aspects
have been addressed: confinement, monopole condensation, glueball
spectra and baryons \cite{Horowitz:1998bj,gross,oz}.
The resulting strong coupling
behaviour of the gauge theories is consistent with
qualitative expectations. As further evidence for the validity of
the gauge/string duality the glueball masses in the supergravity
approximation have been computed \cite{glue1,glue2,glue2vis} with
a fairly good quantitative agreement with lattice data \cite{lat}.
However, there still remain some drawbacks to be clarified, like
the presence of Kaluza--Klein (KK) particles with masses of the
order of the glueballs \cite{glue3}. These states should be absent
in the Yang--Mills theory but the supergravity description
includes them. To overcome this problem, other models for
large-$N$ QCD, generalising \cite{witten} with the use of extra
parameters, have been proposed \cite{Russo:1998mm} and
investigated \cite{Csaki,glue4,Csaki:1999vb}. It turns out that
although it is possible to decouple the KK states associated with
the $S^1$ direction, other unwanted KK states remain coupled.

Models with less than maximal supersymmetry have also been
constructed. Let us simply mention for ${\cal N} =1$ the Klebanov--Strassler
(KS) model \cite{Klebanov:2000hb}, relying on previous work in
\cite{Klebanov:1998hh} and \cite{Klebanov:2000nc}, and that of
Maldacena--N\'u\~nez (MN) \cite{mn}, based on a supergravity solution found in
\cite{Chamseddine:1997nm}. Both models have similarities, lead to
confinement, include chiral symmetry breaking, and are supposed to
describe the same SYM theory in the IR. In \cite{Caceres,Krasnitz}
the glueball spectra for the KS model has been partially studied
although, as far as we know, the KK states and the issue of their
possible decoupling has not been addressed. With regard to these
models, in this paper we shall focus our attention on the
second, of which we shall compute glueball masses as well as
discuss the decoupling of KK states. This is not by far a
straightforward affair: the reason why problems are to be
expected is that the metric is no longer asymptotically
AdS-like, and the conformal symmetry of the theory is explicitly
lost. This is reflected in the running of the dilaton field that
spoils the supergravity approximation at the ultraviolet (UV)
limit. At high energies the theory does not remain a
proper local 4-dimensional QFT, but flows to a little string
theory that will not allow to define sensible two-point functions.

The computation of glueball and KK spectrum in the supergravity
side is closely linked to the existence of confinement. This last
feature is exhibited by an area law behaviour of the Wilson loop
operator and constitutes part of the tests that the supergravity
approximation must fulfil in order to provide us with a reliable
dual description of a confining Yang--Mills theory. In order to
compute the correlation function for the QCD operator using the
gauge/string duality we shall need to consider the dynamics of the
small fluctuations of the supergravity fields \cite{conjecture2}:
{f}or any string mode, $\varphi$, there must exist its
corresponding gauge invariant operator in the field theory, ${\cal
O}$, to which it couples at the boundary with a Schwinger
term $W[\varphi] \equiv \int\, d^4x \varphi(x)
{\cal O}(x)$
\begin{equation}
\label{corre}
\Big{\langle} e^{-\int\, d^4x \varphi_0(x) {\cal O}(x)}
\Big{\rangle} =e^{-S_{\rm SG}[\varphi_0(x)]}\,,
\end{equation}
with $S_{\rm SG}$ the supergravity action and $\varphi_0$ the
value of the field at the boundary $\varphi_0(x) =
\underset{\rho\rightarrow\infty}{\lim} \varphi(x,\rho)\,.$ More
precisely we are interested in scalar operators of QCD: we shall
compute correlation functions in the r.h.s. of \bref{corre} at
large-N and large 't Hooft  coupling constant $g^2 N$, where the
supergravity approximation is valid, and look for particle poles.
Following \cite{witten}, \cite{oz} we shall assume that the
dilaton field couples to the operator ${\cal O}_4 \propto {\rm Tr}
F^2$ in the l.h.s of \bref{corre}, although there can be a certain
mixing with other scalars like the axion
field. Expanding \bref{corre} and in the absence of fluctuations
for the gravitational, RR and/or NS--NS fields in type--IIB string
the equation of motion (e.o.m.) of the scalar fluctuations
$\tilde{\Phi}$, is given in the string frame by the Laplace
operator
\begin{equation}
\label{laplace}
\partial_\mu\left( \sqrt{g} e^{-2 \Phi} g^{\mu \nu} \partial_\nu
\tilde\Phi \right) = 0\,.
\end{equation}
This expression constitutes the simplified basis of our findings.

As mentioned the calculations that we shall perform in this work
are expected to be valid for SU($N$) gauge theories in the
large-$N$ limit and for large 't Hooft coupling constant $g^2 N$.
In this special limit it is known that supergravity is a valid
approximation. Surprisingly enough, our results for QCD$_3$ would
lie within present-day lattice data that are taken just in
the opposite corner, at small 't Hooft coupling constant.

The content of this paper concentrates on two backgrounds and
their respective non-commutative extensions.
The inclusion of non-commutative effects does not change the
pattern of decoupling of the KK states. In full generality, the
non-commutative parameter $\theta$ acts as the electric potential,
$V_{\rm e}(x) = -{\cal E} x$, in the cool emission of electrons
from a metal. Within a metal the electrons are bounded and when
the metal is in its lowest energy state all the energy levels are
filled up to the Fermi level, $E_F$. The minimum energy to remove
an electron from the metal is the \emph{work function}, $W$: the
difference between the top of the well, $V_0$, and the Fermi
energy. In switching on the electric field (at zero temperature)
the binding potential changes accordingly to $V(x) = V_0 - {\cal
E} x$ and the transmission coefficient is given
by the Fowler--Nordheim expression
$$
T = exp\left(-\frac{4}{3} \frac{(2 q m)^{1/2} W^{3/2}}{{\cal
E}}\right)\,.
$$
In increasing ${\cal E}$ the transmission coefficient grows up to
1: more and more less energetic electrons can escape from the
metal and the number of energy levels decreases.

In section
\ref{n=0} we shall introduce our notation. {F}or this purpose we
study the non-commutative version of the model presented in
\cite{witten} for QCD$_3$ \cite{maldajorge}. We shall discuss our
findings, glueball masses and $S^5$ KK states in terms of the WKB
approximation.
Section \ref{n=1} introduces the basics of the MN model followed
by a discussion on the supergravity spectra. The calculation of
the glueball spectra is by far nontrivial in this case: we first
disentangle the main problem, the running away of the dilaton.
Introducing a hard cut-off, we
obtain the glueball spectra.
We then turn to the discussion of the $1^{--}$
states. The role of the non-singlet KK states is discussed in
the last subsection \ref{kk}. Our conclusions and findings are
collected in the Summary. Finally some comments
concerning the WKB methods are gathered in the appendices.

\setcounter{equation}{0}
\section{Setting up Notation: non-commutative QCD$_3$}
\label{n=0}

To begin with we consider a non-supersymmetric model of QCD in
$3$-dimensions \cite{witten}. Our aim is twofold: firstly, to
settle the notation and the main lines of argumentation we shall
follow; and secondly, the model will constitute the playground for
the next and more involved model to be analysed, where we shall
see that much of the basic patterns concerning non-commutativity
are common to the QCD$_3$ model. In order to exemplify the
techniques we restrict our attention to the non-commutative version
of \cite{witten} as described in \cite{maldajorge}. The spectra of
the commutative case has been already scrutinised carefully in
several places \cite{glue1,glue2,glue3,glue4}. In principle
non-commutative effects seem to be associated with high-energy
processes and one should wonder of their relevance at the infrared
scale, for instance the impact in the glueball spectra.
There are essentially two main reasons to keep track of the
effects of non-commutativity. The first one is given by the fact
that as mentioned in \cite{maldajorge} the effects are
unexpectedly of large range in this precise model. And secondly, as
is well-known the commutative version of the model has no mass gap
in the spectra between the singlet states and their corresponding
Kaluza--Klein \cite{glue3}. Many efforts have been made to
circumvent this last drawback, at the expenses of increasing the
number of parameters in the model \cite{Russo:1998mm}, but all
have been so far unsuccessful \cite{Csaki:1999vb}. That is why we
find still worthy to search for any other sources that can lead to
the desired property of decoupling between the singlet states and
the non-singlet KK ones.

The model we shall consider is constructed by compactifying the
Euclidean time direction of a non-extremal D3-brane geometry that
in the sequel should play the role of an internal coordinate.
Afterwards, in order to obtain the ``field-theory'' limit, one has
to rescale the transverse coordinate by sending $\alpha^\prime\rightarrow 0$.

The non-commutative version of the metric requires in addition some
further steps. These are, roughly speaking: ${\it i})$ a
chain of T-dualities, in our concrete case, on the $x_2$ and $x_3$
directions. These operations delocalise the brane along
the $(x^2,x^3)$ plane. ${\it ii})$ Then we turn on a
constant $B_{23}$-field, still keeping the equations
of motion satisfied. And finally, ${\it iii})$ T-dualising back
on the $x_2$ and $x_3$ directions. This last step mixes the
RR-field with the non-commutative parameter $\theta$ in a
nontrivial manner. The solution in the string metric is given,
after a Wick rotation to Euclidean signature, by \cite{maldajorge}
\footnote{Henceforth we shall work in string units.}
\begin{equation}
\label{qcd}
ds^2_{\rm str}=R^2 \left[
 h(u)
d\tau^2 + u^2 dx_1^2 +\frac{u^2}{1+\t^4 u^4}(dx_2^2+dx_3^2)
+h(u)^{-1}
du^2 + d\Omega_5^2 \right]\,,
\end{equation}
with
\begin{equation}
R^4 = 4\pi N\,, \quad h(u)=  u^2 - \frac{u_H^4}{u^2}\,.
\end{equation}
Besides the obvious changes in the warp function that affect
the non-commutative directions there are changes in the
supergravity fields. The only one relevant for us
is the dilaton, which reads
\begin{equation}
e^{2\Phi}= \left( 1+\t^4 u^4\right)^{-1}\,.
\end{equation}
If we set to zero the non-commutative parameter we recover smoothly
the usual background metric and dilaton.

In order to solve \bref{laplace} and calculate the spectrum of
masses we chose a static plane wave ansatz along the
$\mathbb{R}^3$ directions,
$\tilde\Phi(u,x_i)=\phi(u) e^{ik\cdot x}$.
Using \bref{qcd} the e.o.m. for the massless dilaton boils down to
\begin{equation}
\label{eq1}
\partial_u\left(u^3 h(u) \partial_u \phi \right)+
u \left( M^2 - \t^4  k_\perp^2 u^4
\right) \phi = 0\,, \quad k_\perp^2\equiv k_2^2+k_3^2\,,
\end{equation}
where we have defined, after going from Euclidean to Minkowskian
in $x_1$, $M^2 \equiv k_1^2 - k_2^2 - k_3^2$. Notice
that the asymptotic behaviour of \bref{eq1} differs from the
commutative parallel case \cite{witten}. Now the asymptotic
behaviour for large $u$ is of the type $\phi \sim e^{-\sqrt{\t^4  k_\perp^2}
u}$. Paraphrasing the language of quantum mechanics, this
indicates that the presence of the non-commutativity parameter $\t$
makes the large $u$ region classically inaccessible, thus
signaling the existence of an upper finite turning point
---separating the classically accessible region from the
inaccessible one--- when equation \bref{eq1} is written in a
Schr\"odinger-like form, as we shall do below.

As in \cite{witten}, the existence of normalisable solutions for
\bref{eq1} at the horizon and at the deep UV should pin down a
discrete set of values of $M^2$ that will constitute the glueball
spectra.
The regularity of the wave function at the horizon imposes the constraint
$$ \frac{d\phi}{du}
\Big\vert_{u= u_H}=0\,,
$$
which in turn implies in this particular case
$
\phi(u_H) =0\,. $

To obtain the spectrum we shall use the WKB approximation, which is
known to give rather accurate results in the commutative case
\cite{Russo1}. We write \bref{eq1} in the more generic form
\begin{equation}
\label{eq1bis}
\partial_u\left( f(u) \partial_u \phi \right) + \left( M^2 q(u) + p(u) \right)
\phi =0\,.
\end{equation}
Notice that \bref{eq1bis} can be read from the point of view of
field theory as the starting point for a momentum dependent
renormalisation \cite{maldajorge}. One can cast \bref{eq1bis} as a
Schr\"odinger-type equation with zero energy
\begin{equation}
\label{sch}
\partial_u^2\psi(u) -V(u)\psi(u) = 0
\end{equation}
after having performed the functional change
\begin{equation}
\label{funchange}
\psi(u) =
f(u)^{1/2} \phi(u)\,.
\end{equation}
This leads to the potential
\begin{equation}
\label{pot1} V(u) =\frac{1}{2} f(u)^{-\frac{1}{2}}\, \partial_u
\left(f(u)^{-\frac{1}{2}}
 \partial_u f(u) \right)
- \left( q(u) M^2+p(u)\right) f(u)^{-1}\,.
\end{equation}

It is often convenient to perform an additional change of variables. Denoting
it in full generality as $u=F(\omega)$, the new potential
entering the Schr\"odinger equation written in terms of $\omega$
is worked out in the Appendix and reads
\begin{equation}
\label{lastpot}
\tilde{V}(\omega) =
V\left(F(\omega)\right)(\partial_\omega F)^2 - \frac{1}{2} Q_F(\omega)\,.
 \end{equation}
Indeed it acts as the usual coordinate change, with a factor due
to the Jacobian, plus a term with the Schwarzian derivative of
$F$, $Q_F(\omega)$.

In order to solve eq.(\ref{eq1}), we have found convenient to define
the variable
 \cite{Russo1} $\omega=
\log\left(u^2-u_H^2\right)$, that is, $u =
(u_H^2+e^\omega)^{\frac{1}{2}}$,
and working in units $u_H = 1$ the potential becomes, in our
case of interest,
\begin{equation}
\label{potqcd3}
\tilde{V}(\omega) = -M^2 \frac{x}{4(1+x)(2+x)}+
\frac{x\left(12+25 x +18 x^2 + 4 x^3\right)}
{4(1+x)^2 (2+x)^2}+\alpha \frac{x (1+x)}{4(2+x)}\,, \quad x
\equiv e^\omega\,,
\end{equation}
where we have defined $\alpha \equiv\theta^4 k_\perp^2$.

\begin{figure}[ht]
\begin{center}
\includegraphics[width=9cm,height=7cm]{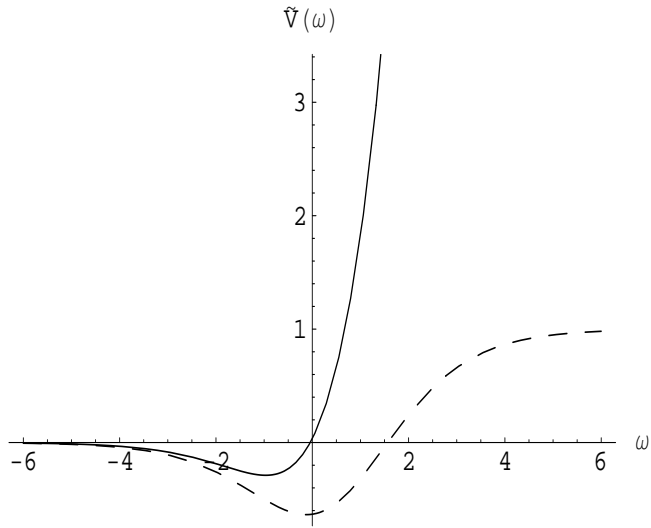}
\end{center}
\caption{\label{fig1}
The potential $\tilde{V}(\omega)$, \bref{potqcd3}, as function of
$\omega$ for two values
of the non-commutative parameter $\alpha$. The solid line corresponds to
$\alpha=4$ while the dashed to $\alpha=0$. For illustrative purposes
we have set
$M=5$.}
\end{figure}

Before analysing the states given by \bref{potqcd3}, it is worth
looking at the qualitative effects of the non-commutative
parameter. We have displayed in fig. \ref{fig1} two different
curves corresponding to two well-separated values of $\alpha$. The
dashed curve corresponds to $\alpha=0$ whilst the full one
corresponds to the case $\alpha=4$. The effect of including
non-commutativity is the shrinking of the area defined by the
negative values of the potential. Therefore, one can intuitively
infer from fig. \ref{fig1} that the mass of the ground state is
``pulled up'' due to the presence of $\alpha \ne 0$. As a matter
of fact the ground state mass
acquires its minimum value for $\alpha=0$. One can understand the
behaviour of the ground state by just comparing with a
1-dimensional square-well potential. In this case the number of
states, N, is given by $(N-1)\frac{\pi}{2} < \gamma \le N
\frac{\pi}{2}$ with $\gamma$ being the strength of the potential,
which in turn is defined, for a square-well potential of length $2
a$ and depth $\vert V_0\vert$, as $\gamma \propto \left(V_0
a^2\right)^2$. Therefore if $\vert V_0\vert$ decreases, as is the case
in fig.~\ref{fig1} when increasing $\alpha$, so does
the number of the bound states.

To substantiate this claim concerning the behaviour of the
ground state mass in relation to the  non-commutativity
parameter, we shall calculate the mass spectrum within the
semi-classical WKB approximation
\begin{equation}
\label{wkb} I(n)\equiv \left( n +\frac{1}{2}\right) \pi =
\int_{\omega_1(M,\alpha)}^{\omega_2(M,\alpha)} d\omega
\sqrt{-\tilde{V}(\omega)}\,, \quad n=0,\, 1,\,2,\, \dots
\,.
\end{equation}
The points $\omega_1$ and $\omega_2$ are the turning points where
$ \tilde V (\omega) = 0$.
Notice that after the
coordinate transformation $u \rightarrow \omega$ the horizon
$u_H = 1$ is located at $\omega_1 \rightarrow -\infty$.

In order to keep track analytically of the contributions to
\bref{wkb} we need an extra approximation to resolve it. {For}
sufficiently large masses $M$ we can expand \bref{wkb} in inverse
powers of $M$. This extra assumption treats the $\alpha$ term as a
small parameter correction and we shall indeed cross-check at
the end the validity of the assumption. The r.h.s of \bref{wkb}
becomes
\begin{equation}
\label{Mexpansion}
I(n) \approx \int_{-\infty}^{\omega_2(M,\alpha)} d\omega \left\{ M \sqrt{A(\omega)}
\left(
1+\frac{1}{2 M^2 A(\omega)} (B(\omega) +\alpha C(\omega) ) \right) \right\}\,,
\end{equation}
where the functions $A,B$ and $C$ can be read directly from
\bref{potqcd3}, $\tilde{V}(\omega) = -M^2 A(\omega) +
B(\omega)+\alpha \, C(\omega)$. Let us start by evaluating the
leading term in \bref{Mexpansion}. Using $x = e^\omega$,
\begin{equation}
\label{approx1}
\int_{-\infty}^{\omega_2(M,\alpha)} d\omega \sqrt{\frac{e^\omega}{(1+e^\omega)
(2+e^\omega)}} \approx \int_0^{\infty} \frac{d x}{\sqrt{x (1+x)(2+x)}}
- \int_{x_2(M,\alpha)}^\infty \frac{dx}{x^{3/2}}\,,
\end{equation}
where in the second term of the r.h.s we have used the asymptotic
expansion of the $A$ function. Both integrals can be performed
analytically in terms of the elliptic functions of the first kind,
$F$, with the result in terms of the leading mass, $
\overset{\circ}{M}$,
\begin{equation}
\label{leading}
(n+\frac{1}{2})\pi = \overset{\circ}{M} \left( F(1\vert 2)
- \frac{1}{\sqrt{x_2( \overset{\circ}{M},\alpha)}}\right)\,.
\end{equation}
Notice that as $\overset{\circ}{M}$ has to be positive defined it is
mandatory that
$
F(1\vert 2) (x_2( \overset{\circ}{M},\alpha))^{1/2} \ge 1$, establishing
a lower bound for $\overset{\circ}{M}$.

The sub-leading term in \bref{Mexpansion} can also be treated
analytically in terms of the elliptic integrals of 1st and 2nd
kind, $F$ and $E$ respectively. The final result is given by
\begin{eqnarray}
\label{massqcd3}
(n+\frac{1}{2})\pi &=& M \left( F(1\vert 2)
- \frac{1}{\sqrt{x_2( M,\alpha)}}\right)
+\frac{1}{4\overset{\circ}{M}}\left\{
\frac{\sqrt{\cx} \left(60+149 \cx +114 \cx^2 +27 \cx^3\right)}
{3 \left[(1+\cx)(2+\cx)\right]^{3/2}}\nonumber \right.\\&&
+ 9 \sqrt{2} \left[ -E\left(\arcsin(\sqrt{2})\vert \frac{1}{2}\right)+
 E\left(\arcsin(\sqrt{2+\cx})\vert \frac{1}{2} \right) \right]
\nonumber \\&& \left.
+ 16 \left[ F\left(\arcsin(\sqrt{-1-\cx})\vert -1\right)-
 F\left(i\, {\rm arcsinh}(1)\vert -1 \right) \right] \frac{}{}\right\}\,,
\end{eqnarray}
with $\cx \equiv x_2(\overset{\circ}{M},\alpha)$. Despite its
apparency \bref{massqcd3} defines a real function. Notice that in
order to solve \bref{massqcd3} we have used two different values
of the masses $M$: the first term, r.h.s., defines the {\sl total}
glueball mass including all the corrections, while up to the order
we are working it will suffice to use the leading correction as
defined in \bref{leading} in the second term of the r.h.s. In that
way we consider the latter as a small correction to the former. This is
indeed verified by the numerical results.

In order to proceed further
all we need
to know is the upper turning point $x_2(M,\alpha)$. For that purpose one
solves \bref{potqcd3} $\tilde{V}(x) = 0$. Certainly we can find analytically
the roots of this equation, but then it will be impossible to invert
\bref{leading}, or equivalently the first term r.h.s. of \bref{massqcd3}.
In turn
the point can be approximated by just keeping up to the sub-sub-leading
corrections in \bref{potqcd3}
\begin{equation}
\label{approx2}
x_{\rm turning} = - \frac{1}{2\alpha}\left( 4+\alpha -\sqrt{\alpha^2
+4 \alpha(M^2-16)+16}\right)\,.
\end{equation}

\begin{table}
\begin{center}
\begin{tabular}{cccc}
\hline
   & {\rm Lattice}, $N=3$ \cite{lat} &
    {\rm Lattice}, $N\rightarrow \infty$ \cite{lat} &
    \CN=1 {\rm WKB} \\
\hline
$0^{++}$ &  $4.329\pm 0.041$  & $4.065\pm 0.055$ & $4.065$ \\
$0^{++*}$ & $6.52\pm 0.09$    & $6.18\pm 0.13$   & $6.6$ \\
$0^{++**}$ & $8.23\pm 0.17$   & $7.99\pm 0.22$   & $9.3$ \\
\hline
\end{tabular}
\end{center}
\caption{\label{table1}
Comparison
of the lowest lying glueball masses between lattice results
and the
prediction based on the QCD$_3$ supergravity inspired model of \cite{witten}.
The ground state has been normalised to the $N\rightarrow \infty$
lattice result.
}
\end{table}

\begin{figure}[t]
\begin{center}
\includegraphics[width=8cm,height=6cm]{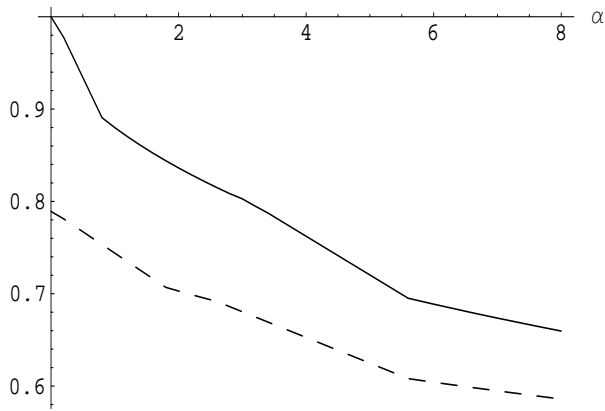}
\end{center}
\caption{\label{fig2}
The ratios $\frac{M_{0^{++}}(\alpha=0,l=0)}{M_{0^{++}}(\alpha,l=0)}$
and  $\frac{M_{0^{++}}(\alpha=0,l=0)}{M_{0^{++}}(\alpha,l=1)}$
(solid and dashed lines respectively)
vs. $\alpha$.
}
\end{figure}

Our main result is collected in fig. \ref{fig2} where we have
plotted the ratio $\frac{M_{0^{++}}(\alpha=0)}{M_{0^{++}}(\alpha)}$ as a
function of $\alpha$ (solid line). Due to the approximations in
\bref{approx1}, \bref{approx2} and the treatment of the mass in
\bref{massqcd3} we should interpret it only qualitatively, but the
result agrees extremely well with the exact one for highly excited
states. A more complete numerical analysis of \bref{wkb} using
\bref{potqcd3} confirms fully the spectra for any principal
quantum number with quite high accuracy. Furthermore, in order to
ascertain the correctness of our results, we have verified that
for $\alpha=0$ we recover the same spectrum found in \cite{glue1}.
For the sake of completeness we depict it in table
\ref{table1}. As it is evident
from fig. \ref{fig2}, the ground state mass increases with
$\alpha$ thus confirming our previous qualitative expectations
derived from the change in the shape of fig. \ref{fig1}. From the
interplay between fig. \ref{fig2} and table \ref{table1}
we notice that the difference between the lattice
results \cite{lat} and the supergravity ones increases with
$\alpha$, because the corrections to the latter go in the opposite
direction. Provided the lattice results are considered robust,
this fact suggest that the allowed room for a value $\alpha \ne 0$
is quite constrained.

\subsection{Kaluza--Klein modes of $S^5$}
\label{kk-qcd3}

 We concluded in the previous section that the inclusion of
a $B$-field in the supergravity picture, accounting for the
non-commutativity in the field theory side, makes heavier the
lowest lying states of QCD$_3$. One should wonder whether this
fact opens the possibility of a decoupling between the singlet
states and the non-singlet ones. As has been already mentioned the
present non-decoupling constitutes one of main drawbacks of the
model. In order to calculate the non-singlet KK masses we do not
project onto the s-wave as before but keep the harmonic spherics
of $S^5$. This slightly changes the previous plane wave ansatz to
$\tilde{\Phi}(u,x_i)=\phi(u) e^{ik\cdot x} Y_l(\Omega_5)$. This will
accordingly modify \bref{eq1} to
\begin{equation}
\label{eq1l}
\partial_u\left(u^3 h(u) \partial_u \phi \right)+
u \left( M^2 - \alpha u^4 - l (l+4) u^2
\right) \phi = 0\,.
\end{equation}
This expression already shows that the role played by the angular
momentum of the $S^5$ and the $\alpha$ parameter is rather
similar. Both have an opposite sign with respect to $M^2$ and
hence any increase in the values of $l$ or in $\alpha$ should lead
to similar qualitative results. Obviously, there are some major
differences between both terms; albeit, while the angular momentum
can affect considerably the infrared behaviour the non-commutative
parameter is dominant in the ultraviolet. A second difference is on the
nature itself of the parameters: the angular momentum terms only
take integer values and then there is a jump between the singlet
state and the first non-singlet KK with $l=1$ while the variation
of the eigenvalues, or masses, with $\alpha$ is a continuous and
smooth function. This suggest that our results are not expected to
deviate significantly from the conclusions derived in \cite{glue3}
concerning the decoupling of the non-singlet KK states provided the
range of the $\alpha$ parameter can not be extended too much,
otherwise non-commutative effects will start to affect the infrared
region, which is not a rather plausible scenario if the lattice
results should be taken for granted.

To substantiate this claim quantitatively we proceed as in the
previous section, obtaining the potential of \bref{potqcd3}
plus an additional term depending on $l$. It explicitly reads
\begin{equation}
\tilde{V}(\omega,l)= {\rm eq.}~\bref{potqcd3} + \frac{x l (l+4)}{4(2+x)}\,.
\end{equation}
Masses are computed in a similar way using the WKB approximation.
The result is displayed in fig.~\ref{fig2} where the
ratio $\frac{M_{0^{++}}(\alpha=0,l=0)}{M_{0^{++}}(\alpha,l=1)}$
is shown in dashed line.
Even if, as expected,
the combined result of the $\alpha$ and $l$ terms increase the
difference between the singlet and the lowest non-singlet KK states,
for reasonable small values of $\alpha$ it is insufficient to
decouple them.

\section{SYM \CN=1: Maldacena--N\'u\~nez model}
\label{n=1}
\setcounter{equation}{0}

Now we would like to apply similar methods as the above presented
in the previous section to obtain the spectrum of glueballs, and
also of non-singlet KK states, for the Maldacena-N\'u\~nez
supergravity background, whose IR regime is dual to a ${\cal N}=1$
SYM theory.

The background is that given by a stack of D5-branes wrapping an
$S^2$ within a Calabi--Yau three-fold.
In the string frame the relevant fields for the Maldacena-N\'u\~nez
background are given by the metric
\begin{equation}
\label{metric} ds^2_{10}=e^{\Phi_D} \left[ dx^2_{0,3}+ N
\left(d\rho^2+e^{2g(\rho)} d\Omega_2^2 +\frac{1}{4}
\sum_{a=1}^3(\omega^a-A^a)^2\right) \right]\,,
\end{equation}
the dilaton field
\begin{equation}
\label{dil-D5} e^{2\Phi_D(\rho)}=
e^{2\Phi_{D,0}}\frac{\sinh2\rho}{2e^{g(\rho)}} \,,
\end{equation}
and the Ramond-Ramond three-form
\be
\label{F3}
F_{[3]}=dC_{[2]}=\frac{N
g_s}{4}\left[-(w^1-A^1)\wedge(w^2-A^2)\wedge(w^3-A^3) +
\sum_{a=1}^3 F^a\wedge(w^a-A^a)\right]\,.
\ee
We shall also introduce the $SU(2)_L$ modified gauge fields
\cite{Bertolini:2002yr}
\begin{eqnarray}
A^{\prime 1}&&\hspace{-0.5cm}
=\frac{1}{2} (a(\rho)-1)\left(-\cos\varphi d\theta +
\cos\theta \sin\theta \sin\varphi d\varphi\right)\,,\nonumber\\
A^{\prime 2}&&\hspace{-0.5cm}
=\frac{1}{2} (a(\rho)-1)\left(\sin\varphi d\theta +
\cos\theta \sin\theta \cos\varphi d\varphi\right)\,,\nonumber\\
A^{\prime 3}&&\hspace{-0.5cm}
=\frac{1}{2} (a(\rho)-1)
\sin^2\theta d\varphi\,,
\end{eqnarray}
related to the ones originally used in \cite{mn} by a
gauge transformation
\begin{equation}
A\rightarrow A^\prime = g^{-1} A g + i g^{-1} d g\,,
\end{equation}
with $g$ being an element of the SU(2) group chosen to be
\begin{equation}
g(\theta,\varphi) = e^{-\frac{i}{2}\theta\sigma_1}
e^{-\frac{i}{2}\varphi\sigma_3}\,.
\end{equation}
With this choice of the expression for the $SU(2)$ gauge
potential, there is a simple identification
of the 2-cycle wrapped by the D5-brane,
in such a way that the 2-cycle no longer mixes with the 3-cycle.
The three-sphere is parameterised by the left-invariant one-forms $w^i$.

In addition we have the following definitions
\begin{equation}
e^{2g(\rho)}\equiv \rho \coth 2\rho - \frac{\rho^2}{\sinh^2 2\rho}
-\frac{1}{4}\,, \quad a(\rho)\equiv \frac{2\rho}{\sinh 2\rho}\,.
\end{equation}

Notice that \bref{metric} is only strictly valid in the infrared region
where the dilaton field remains bounded and hence supergravity is valid. If
one wishes to push the theory to the ultraviolet one has to perform an
S-duality obtaining the metric
\begin{equation}
\label{smetric} ds^2_{10}= dx^2_{0,3}+N \left(d\rho^2+e^{2g(\rho)}
d\Omega_2^2 +\frac{1}{4} \sum_{a=1}^3(\omega^a-A^a)^2\right)\,,
\end{equation}
and the dilaton
\begin{equation}
\label{dil-N5}
e^{2\Phi(\rho)}=e^{2\Phi_{NS,0}}\frac{2e^{g(\rho)}}{\sinh2\rho}\,.
\end{equation}

\subsection{SUGRA spectra: preliminaries}

To obtain the spectra of the theory
in full generality, instead of the duality argument that lead to
the use of \bref{laplace}, one should consider the linearised e.o.m.
of the bosonic sector in type-IIB string theory in a given background
\cite{joe}.
The variation of the e.o.m. for the dilaton and the RR 3-form
considering only linear fluctuations give \cite{Schwarz:qr}
\begin{eqnarray}
\label{realeom}
D^{\hat{\mu}} \partial_{\hat{\mu}} \tilde{\Phi} = \frac{\kappa_{10}^2}{24}
\dot{g}_{\hat{\rho} \hat{\sigma} \hat{\tau}}
g^{\hat{\rho} \hat{\sigma} \hat{\tau}}\,, \quad
D^{\hat{\mu}} g_{\hat{\mu} \hat{\rho} \hat{\sigma} } = -\frac{2i \kappa_{10}^2}
{3} F_{\hat{\rho} \hat{\sigma} \hat{\tau}\hat{\lambda}\hat{\nu}}
\dot{g}^{ \hat{\tau}\hat{\lambda}\hat{\nu}} + \partial^{\hat{\mu}}
\tilde{\Phi}
\dot{g}_{\hat{\mu} \hat{\rho} \hat{\sigma}}\,,
\end{eqnarray}
where we have closely followed the notation of \cite{Kim:ez}
\footnote{ Caret indices stand for 10-d coordinates, Latin for
CY$_3$ variables and Greek for the non-compact 4-d space
coordinates. Dotted fields refer to the classical ones, un-dotted
to the fluctuations. $D$ stands for the covariant derivative.}. In
principle one has to project each of the bosonic fields on the
CY$_3$ manifold by the proper spherical harmonic and hence reduce
the fields to the non-compact space in 4-d. In practise this is
cumbersome and some simplification is needed. The Calabi--Yau
three-fold contains the product $S^2 \times S^3$. The roles of both
$S^d$ are very different, while the $S^2$ shrinks to zero in the
IR the $S^3$ still remains with a nonzero radii at $\rho=0$. This
indicates that low-energy dynamics should be ridden by the effects
of the $S^2$, while the effects of the $S^3$ are of higher energy
and thus can be decoupled in this region. As a consequence the
solutions for the fluctuations can be expanded in
spherical harmonics on the $S^3$ instead of the harmonics of the
exact background. Following \cite{Kim:ez} we decompose the
fluctuations of the scalar field \bref{dil-D5} and those of
the complex two-form (formed as a combination of
the RR and NS--NS potentials) by projecting on 
$\mathbb{R}^4 \times \mathbb{R} \times S^2$ with the help of  
hyper-spherical harmonics on $S^3$. 

\subsection{Glueball masses: boundary conditions}

To begin with we are interested in the lowest physical state: i.e.
the scalar state or glueball mass.
Matching the $S^3$ spherical harmonics of both sides of
the first expression of \bref{realeom} one recovers \bref{laplace}
for the s-wave mode. As previously we shall assume
a static mode solution, $x_i\in \mathbb{R}^4 \times \mathbb{R} \times S^2$,
\begin{equation}
\label{ant}
\Phi(x_i,\rho,\Omega^2)=\tilde{\Phi}(\rho) e^{ik\cdot x} Y^m_l(\Omega^2)\,,
\end{equation}
Setting $l=m \equiv 0$ and replacing the metric by \bref{metric},
the equation for the fluctuations of the massless dilaton
\bref{laplace} takes the form
\begin{equation}\label{dil-MN}
\partial_\rho(e^{2\Phi_D(\rho)}e^{2g(\rho)}\partial_\rho
\tilde{\Phi})+ N M^2 e^{2\Phi_D(\rho)}e^{2g(\rho)}
\tilde{\Phi}=0\,.
\end{equation}
Notice that the dilaton e.o.m. for the S-dual metric
\bref{smetric} is exactly the same expression
\bref{dil-MN}. The reason for this coincidence is that both 
equations correspond to the Laplacian equation in the Einstein 
frame, and the metric in the Einstein frame is invariant under S-duality.

\begin{figure}[t]
\begin{center}
\includegraphics[width=16cm,height=1cm]{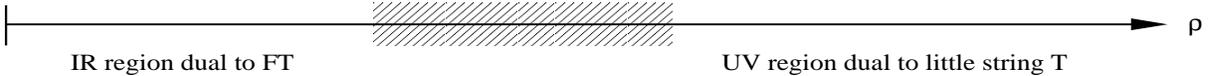}
\end{center}
\caption{\label{phs} The phase structure in the MN model. Only the
IR region is dual to a field theory. In the intermediate region
neither \bref{metric} nor its S-dual \bref{smetric} apply. }
\end{figure}

Let us mention that the gauge fields $A^a$, present in the metric
\bref{metric} do not contribute to the determinant of the
metric. As before $ M^2 = k_0^2-\vert \vec{k}\vert^2$, but without
need of performing any Wick rotation. Since at the end the value
of $M$ should be normalised to the ground state one, in the
following we shall rename the coefficient $N  M^2$ as just $M^2$.
Before proceeding by writing \bref{dil-MN} as a Schr\"odinger-like
equation let us inspect it carefully. In ordinary quantum
mechanics an expression similar to \bref{dil-MN} does not describe
a discrete set of bound states unless some requirements are
satisfied. First, the fact of being a discrete spectra relies
entirely on the boundary conditions (BC): the solutions of
\bref{dil-MN} should properly match one BC condition at the origin
of the radial coordinate and another at the boundary. These pin
down the precisely allowed values of $M^2$. And second, in order
to describe a bound state system, at least one of the solutions of
\bref{dil-MN} should be normalisable with some measure
\begin{equation}
\label{norm0}
\int_0^\infty d\rho\sqrt{g}\, \vert \tilde{\Phi} \vert^2\,,
\end{equation}
being $g\equiv g_{\rm IR},g_{\rm UV}$ the determinant of the
metric \bref{metric} and \bref{smetric} respectively. To deal with
the normalisation of the wave function we shall bear in mind the
phase structure of the theory. At low-energy, where the model is
meant to be dual to ${\cal N} =1$ SYM field theory, the relevant
fields are \bref{metric} and \bref{dil-D5}. In increasing the
energy (assumed to be given by the radial coordinate) the scalar
curvature becomes bigger and the sugra approximation breaks down,
hence one has to deal with the S-dual metric \bref{smetric} and
the dilaton \bref{dil-N5}. Then the normalisation of the wave function 
\bref{norm0} should be replaced by the expression
\begin{equation}
\label{norm} \int_0^\infty d\rho \sqrt{g}\, \vert \tilde{\Phi}
\vert^2 \rightarrow \int_0^{\Lambda} d\rho \sqrt{g_{{\rm IR}}}\,
\vert \tilde{\Phi} \vert^2 + \int_{\Lambda}^\infty d\rho
\sqrt{g_{{\rm UV}}}\, \vert \tilde{\Phi} \vert^2 = 1\,,
\end{equation}
with $\Lambda$ a hard cut-off scale denoting the range of applicability
of both descriptions (see below).
{F}or simplicity we have discarded an intermediate region
where neither \bref{metric} nor \bref{smetric} are applicable, see
fig.~\ref{phs}. At the level of the dilaton e.o.m. \bref{norm} has to be
interpreted in the following way: in principle one has to solve \emph{two}
differential equations (corresponding each one to one of the two terms 
in r.h.s. \bref{norm} )
that in our precise case reduce to a single one. In the IR expression
one has to demand a BC at $\rho=0$ while at the opposite edge one has to deal
with the second differential equation and the BC at $\rho \rightarrow \infty$.
One has to pin down those solutions to the previous system that match smoothly
in the intermediate region, around $\rho \sim \Lambda$. As we shall see
this is unnecessary in our case.

We analyse first the IR behaviour of \bref{dil-MN} where the dilaton field
is given by \bref{dil-D5}. At the origin
of the radial coordinate, $\rho=0$, the wave function is no
longer a smooth function unless we require some extra condition.
In fact choosing a 10-d static-like ansatz as \bref{ant} formally
constrains the form of the wave function. {F}or a fixed point with
coordinates $(x^\mu,\rho)$ on $\mathbb{R}^4\times \mathbb{R}$,
$\tilde{\Phi}$ moves on the $S^2$ and $S^3$ sphere with a constant value
and hence the wave function is stationary at $\rho=0$,
\begin{equation}
\label{bc2}
\frac{d\tilde{\Phi}(\rho)}{d\rho}\Big{\vert}_{\rho=0} = 0\,.
\end{equation}
This can also be seen by direct computation, obtaining the
solution of \bref{dil-MN} near the origin, similarly as was done
in \cite{glue2}. This leads to an even function,
$\tilde{\Phi}(\rho) = \sum a_{2n} \rho^{2n}\,, \quad a_{2n}
\propto a_0$, that fulfils \bref{bc2}.

\begin{figure}[t]
\begin{center}
\includegraphics[width=8cm,height=8cm]{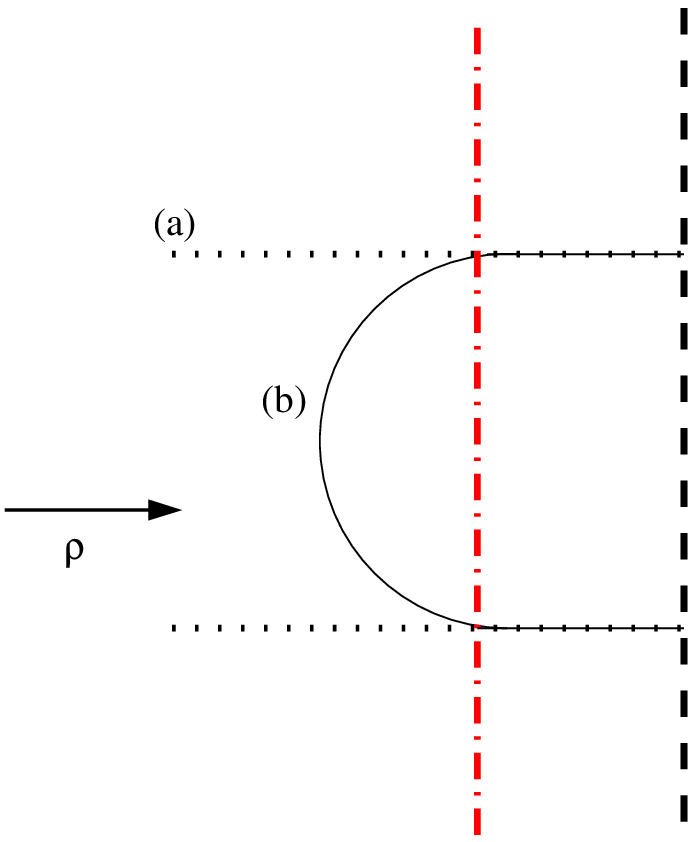}
\end{center}
\hspace{9cm} $\rho=\Lambda_{{\cal N}=1}$ \hspace{0.5cm} $\rho =
\infty$ \caption{\label{brane} Different configurations for the
open string world-sheet in the evaluation of the Wilson loop. (a)
corresponds to the subtraction of ``two bare quarks''. (b) is the
only allowed configuration (fine tuned) that leads to finite results
after renormalisation, for both the potential and the quarks
distance. Notice that configuration (b) hits the brane at right
angles and, therefore, tends asymptotically to (a) for large
$\rho$. }
\end{figure}

A second boundary condition is still lacking. This is settled at
the boundary of the radial coordinate $\rho\rightarrow\infty$,
where \bref{metric} is no longer valid and has to be changed by
its S-dual version \bref{smetric}. At large $\rho$ the behaviour
of \bref{dil-MN} becomes (keeping leading and sub-leading terms only)
\begin{equation}
\partial^2_\rho \tilde{\Phi}(\rho) +
\left(2 +\frac{1}{2 \rho}\right) \partial_\rho \tilde{\Phi}(\rho)+
M^2 \tilde{\Phi}(\rho)=0\,,
\end{equation}
whose solution can be cast in terms of
the confluent hypergeometric functions $_1F_1$ and $U$
\begin{eqnarray}
\tilde{\Phi}(\rho) &\sim & e^{-\left(1+\sqrt{1-M^2}\right) \rho} \sqrt{\rho}\,
\left[A(1)\,\,
_1F_1(\frac{3}{4}+\frac{1}{4\sqrt{1-M^2}},\frac{3}{2};2 \sqrt{1-M^2} \rho)
\nonumber \right.\\&& \left.
+ A(2) \,\,
U(\frac{3}{4}+\frac{1}{4\sqrt{1-M^2}},\frac{3}{2},2 \sqrt{1-M^2} \rho)
\right]\,.
\end{eqnarray}
Both solutions are physically acceptable because they are
normalisable according to \bref{norm}. At this point, and following the
arguments given in \cite{witten} we can conclude that there
is no discrete spectrum (we will give complementary arguments in
the next subsection). Nevertheless, notice that in moving to the S-dual
sugra solution for the UV region, we are no longer describing a
SYM gauge field theory, but a little string theory. The
description of the SYM theory is restricted to the IR region, with
a sugra solution describing D5-branes wrapping an $S^2$. In view
of these considerations, an alternative way to proceed is then to
implement a hard cut-off at some scale $\rho \sim \Lambda$ that
will essentially delimit the IR region. With such an approach, the
normalisability condition will become
\begin{equation}
\label{norm-IR} \int_0^\infty d\rho \sqrt{g}\, \vert \tilde{\Phi}
\vert^2 \rightarrow \int_0^{\Lambda} d\rho \sqrt{g}\, \vert
\tilde{\Phi} \vert^2 \approx 1\,,
\end{equation}
which in turn implies (in addition to the BC \bref{bc2})
to take from
$\Lambda$ up to the deep UV a zero value for the
dilaton wave function
\begin{equation}
\label{bc}
\tilde{\Phi}(\rho) \vert_{\rho \ge \Lambda} = 0\,.
\end{equation}
Notice that this implies a non-analytic behaviour 
for the wave function 
at $\rho = \Lambda$. This is a
simple way of getting rid of the unwanted UV contribution
in \bref{norm}. This can be
seen a quite \emph{unnatural} and \emph{ad hoc} procedure.
However, we should say that it is perhaps the \emph{simplest} way
of giving some interpretation to the lack of glueball states in
the model as it stands. 
In this new approach, the metric field \bref{metric} is
interpreted \emph{only} as an \emph{effective low-energy field};
as one increases the energy the supergravity approximation breaks
down, the dilaton field is unbounded and \bref{metric} has to be
substituted by its S-dual metric \cite{mn}. This implies that only
a restricted region in the variable $\rho$ can be under
perturbative control and therefore it is doubtful that one can
obtain sensible results from the UV region unless some
regularisation is implemented. This reasoning is implicitly used
in the computation of the Wilson loop \cite{rey,wilson}, which we
briefly review: given a 4-d spacetime $M$ we can take a closed
oriented curve $C\in M$ (the Wilson loop) as the boundary of a
compact oriented surface $D$ that, in minimising its area, will
experience the geometry in the bulk of a 10-d space defined by
\bref{metric}. The area of $D$ is infinite but can be regularised
by a counter-term proportional to $C$ giving $\alpha(D)$. The
expectation value of the Wilson loop is proportional to
$e^{-\alpha(D)}$, with $D$ minimising the functional $\alpha(D)$.
Precisely the area of $D$ is regularised by subtracting ``some UV
contribution'', a counter-term, hence keeping only some finite
low-energy information. {F}or the original formulation of
\cite{wilson} the actual procedure is sketched in
fig.~\ref{brane}: the curve (b), representing the intersection of
$D$ with the equal-time surface, has a divergent length and needs
the subtraction of the ``bare quark masses'', configuration (a).
To cancel properly the divergence it was shown in \cite{nc} that
configuration (b) should approach asymptotically (a) and both
should hit orthogonally the brane at $\rho\rightarrow\infty$. The
point addressed in the previous discussion is concerning on the
exact point the cancellation should start to occur. We firmly
believe that to make sense of the model this should start to
happen at quite low $\rho$ (see fig.~\ref{brane} and below) this
is so because, contrary to the Wilson loop calculation, we have not
implemented any procedure of regularisation in \bref{dil-MN}.

Before proceeding let us comment on the possible choices of this
scale and its role. Admittedly, any choice seems to be arbitrary to some extent
and one should wonder on a reliable value of $\Lambda$ that
delimits the IR region. This in turn will have a direct impact on
the uncertainty of the results. To be more definite let us remind
the structure of the phases in the theory flowing from the UV to
the IR region, fig.~\ref{phs}: in the UV region
the theory is R-symmetric and the dynamics is ridden by the fields
\bref{smetric} and \bref{dil-N5}.
In decreasing the energy the value of $\Lambda_{{\cal N}=1}$ is
settled by the point $\rho^*$ where the gaugino starts to
condensate, i.e. the R-symmetry is broken. Correspondingly the
order parameter in the supergravity side is the function $a(\rho)$
in \bref{metric} \cite{cotrone} and $\Lambda_{{\cal N}=1}$  is
given by the point where $a(\rho^*)$ is sensibly different from
zero. The shape of the function $a(\rho)$ is
rather steep, decreasing rapidly from $a(0)=1$ to the zero value
when $\rho$ increases. Indeed for $\rho \approx 5$, $a(5) \approx
10^{-3}$. Albeit we neither claim that this is the precise point nor
that it has something special, we can safely say that we are already
leaving the IR region and that the R-symmetry is in process of
being restored. Hence \CN=1 has an intrinsic scale and we
shall discuss the two possible choices of the $\Lambda$ scale used
in \bref{norm} with respect to it:

\avall

{\it i)} $\Lambda > \Lambda_{{\cal N}=1}$. This implies that the
original model describes at least two different phases in the theory. This
could be signaled by a discontinuity
in some quantity as the Wilson loop. So far this has not
been observed \cite{mn,nc}.

\avall

{\it ii)} $\Lambda < \Lambda_{{\cal N}=1}$. The full theory remains in a single
phase. It is known from effective field theory that in the intermediate
energy range $\Lambda < E < \Lambda_{{\cal N}=1}$ there is no new threshold
for massless states \cite{Appelquist:1992ft} whilst massive modes have been
integrated in our case.

\avall

Therefore without any loss of generality we can identify
$\Lambda \approx \Lambda_{{\cal N}=1}$ in the remainder.
One should wonder of the different treatment of \bref{norm}
in \cite{Caceres} and the absence of the intrinsic scale
$\Lambda_{{\cal N}=1}$ in their argument. Essentially
the two supergravity models are expected to be identical in the IR region, thus
the difference should come from the treatment of the UV contribution.
In \cite{Caceres} \bref{norm} is not splitted into two pieces. This is just as
a consequence of the warp factor appearing in the conifold metric which decays
extremely fast, hence catching all its contribution from the IR region. And
none of the fields present in that model grows unbounded. In other
words the second term in the r.h.s. of \bref{norm} has, if any, almost a
negligible contribution. This is not the case if one insists in using 
only the IR metric \bref{metric} in the full range of $\rho$ for MN.

\subsection{Glueball masses for \CN=1?}

As we have just discussed, albeit we can find a normalised solution
for the dilaton wave function there is at least some problem in choosing
the correct UV behaviour. To substantiate this point
we can proceed analogously
as in the QCD$_3$ case and obtain the associated potential.
Starting from \bref{dil-MN}
and following the notation of \bref{eq1bis}
we identify the functions
\begin{eqnarray}
f(\rho) &=& e^{2\Phi_D(\rho)}e^{2g(\rho)} = 
\frac{e^{2\Phi_{D,0}}}{2} ({\rho
\coth (2\rho) - \frac{\rho^2}{\sinh^2 (2\rho)}
-\frac{1}{4}})^{\frac{1}{2}} \sinh (2\rho)\,, \\ \nonumber
q(\rho)&=& M^2 f(\rho)\,,\quad p(\rho) \equiv 0\,,
\end{eqnarray}
which allow to write a Schr\"odinger-like equation \bref{sch}
for $\Psi = f^{1/2} \tilde{\Phi}$, with the potential \bref{pot1} given by
\begin{figure}
\begin{center}
\includegraphics[width=9cm,height=7cm]{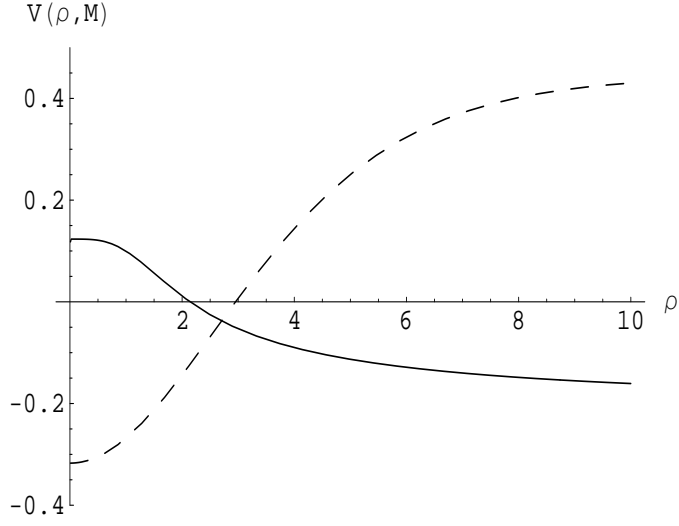}
\end{center}
\caption{\label{figMNoriginal}
The potential $V(\rho,M)$ for the MN model in its original variable
$\rho$ (full curve) and the KS potential (dashed curve),
for a value $M=\frac{4}{3}$.}
\end{figure}

\begin{eqnarray}
\label{MNpot-original}
V(\rho,M) &&\equiv -M^2 + \tilde{V}(\rho)\,,\nonumber \\
\tilde{V}(\rho) &&= -
\frac{2\,\left( -1 - 20\,\rho^2 + \left( 1 - 4\,\rho^2 \right)
\,\cosh (4\,\rho) +
       32\,\rho^3\,\coth (2\,\rho) \right) \,
     {{{\rm csch}^2}(2\,\rho)}}{{\left( 1 -
        4\,\rho\,\coth (2\,\rho) +
        4\,\rho^2\,{{{\rm  csch}^2}(2\,\rho)} \right) }^2} \,.\nonumber \\
\end{eqnarray}
Notice that this expression is valid in the IR or in the UV region.
The form of the $V(\rho,M)$ potential is rather
peculiar and
allows us to rewrite \bref{sch} in terms of the reduced potential
$\tilde{V}(\rho)$ ($M$ independent) as
\begin{equation}
\label{sch2}
-\partial^2_\rho \Psi(\rho) + \tilde{V}(\rho)
\Psi(\rho) = M^2 \Psi(\rho)\,,
\end{equation}
which can be interpreted as a Schr\"odinger equation with
``energy'' $M^2$, contrary to the QCD$_3$ case where the
analogous equation is an eigenvalue problem only with a
vanishing energy. Despite its analytic appearance,
\bref{MNpot-original} exhibits the simple form depicted in
fig.~\ref{figMNoriginal}. (Together with it we have also displayed
the potential corresponding to the KS model
\cite{Klebanov:2000hb}.)
The potential resembles a {\sl smooth} 1-dimensional
step-potential in quantum mechanics, where it is well known that
there are no bound states, the solutions with
$E > \tilde{V}(\rho\rightarrow\infty)$ are oscillatory at infinity.
The non-normalisability of the wave function
solutions is reflected in the asymptotics of the
potential $\tilde{V}(\rho)$, (see also fig. \ref{figMNoriginal}),
\begin{equation}
\label{potlim}
\lim_{\rho \rightarrow \infty} \tilde{V}(\rho) = 1 \quad
{\rm and} \quad
\lim_{\rho \rightarrow 0} \tilde{V}(\rho) =
\frac{4}{3} \,.
\end{equation}

It is worth noticing at this point the striking difference of
shapes in fig.~\ref{figMNoriginal} between the Schr\"odinger-like
potentials for the MN model and the KS model. In sharp contrast
with the  former, the latter can --and eventually will--
exhibit a discrete spectrum of normalisable wave functions.

As we have advanced in the previous subsection, 
one possible way of obtaining a discrete
glueball spectra is by \emph{imposing} 
the BC \bref{bc2} and \bref{bc} for
the $\tilde\Phi$ function which correspond to
\begin{equation}
\label{noubc}
\Psi(0) = \Psi(\Lambda_{{\cal N}=1}) = 0\,,
\end{equation}
for the $\Psi$ function \footnote{Note that $\Psi(\rho) =
f(\rho)^{1/2}\tilde\Phi$, and that $f(\rho)^{1/2}$, analytically
extended to negative values of the variable $\rho$, is an odd
function. In this precise case the BC at $\rho=0$ have moved
from $\tilde\Phi^\prime(0) = 0$ to $\Psi(0)=0$, with the
supplementary information that now $\Psi^\prime(0) =
\tilde\Phi(0)$.}. After this the potential
\bref{MNpot-original} is supplemented with two infinite walls at
$\rho = 0\,,\, \Lambda_{{\cal N}=1}$. It resembles a
one-dimensional infinite square well problem of width
$\Lambda_{{\cal N}=1}$ and whose bottom is slightly distorted with
a mean value given by the average value of \bref{MNpot-original},
i.e., $V_{\rm{well}} = \langle\tilde{ V}(\rho)\rangle$. Notice
that this approach assumes that whatever sufficient smooth
potential in the region $0<\rho<\Lambda_{{\cal N}=1}$ will provide
similar results. In this simplified approximation, the energy
levels (which correspond to the allowed values of masses squared)
are given by
\begin{equation}
\label{crude}
M^2_{n} = \left( \frac{\pi}{\Lambda_{{\cal N}=1}}\right)^2  n^2
+V_{\rm{well}} \,, \quad  (n=1,2,\ldots)\,.
\end{equation}
They only depend on the well width and on $V_{\rm{well}}$.
Notice that for the special case of $V_{\rm{well}}=0$, the masses,
normalised to the lightest one, are in the fixed ratios
\begin{equation}
\label{sqwell}
\frac{M_n}{M_1} \approx  n \,, \quad  (n=1,2,\ldots)\,.
\end{equation}
In our case $V_{\rm{well}} = \langle\tilde{ V}(\rho)\rangle
\simeq 1$ (see fig.~\ref{figMNoriginal} and \bref{potlim}), and
the mass ratios will depend on the actual value of the cut-off
$\Lambda_{{\cal N}=1}$ (see below).

\begin{table}
\begin{center}
\begin{tabular}{cccccc}
\hline
 {\rm n} & {\rm State}  & {\rm Lattice} \cite{lat} & \CN=0 \cite{glue1} &
\CN=1 {\rm KS} \cite{Caceres,Krasnitz} & \CN=1$^*$ \cite{nick}\\
\hline
$1$ & $0^{++}$ &  $1$    & $1$   & $1$    & $1$\\
$2$ & $0^{++*}$ & $1.75$ & $1.4$ & $1.5$  & $1.5$\\
$3$ & $0^{++**}$ & --    & $1.9$ & $2$    & $1.9$\\
$4$ & $0^{++***}$ & --   & $2.3$ & $2.5$  & -- \\
\hline
\end{tabular}
\end{center}
\vspace{0.25cm}
\caption{\label{comp}
Glueball mass predictions of several supergravity models
and the lattice computation.
The ground state has been normalised
to $1$ in all the cases.
The lattice results are subject to errors.
}
\end{table}

It is instructive, even at this early stage, to compare these
crude estimates with predictions from other models.
The models considered and their corresponding results are shown in
table \ref{comp}, where we have normalised all masses to the
lightest, corresponding to the ground state.
A common feature to the models displayed in table \ref{comp}
is that the glueball spectra follows approximately the relation
\begin{equation}
\label{models}
\frac{M_n}{M_1} \approx \frac{n+1}{2}\,\,(n=1,2,\ldots)\,.
\end{equation}

The similitude
between the models depicted in table~\ref{comp} is not surprising.
Previous calculations suggest
that the ball park value of the
glueball masses is almost independent
of the
supergravity model \cite{glue4} used to calculate them. This can
be understood on the basis of the WKB approximation from
which it follows that the main contribution to the masses
is identified by the leading contribution (the first term, r.h.s. in
\bref{massqcd3} in the QCD$_3$ case) giving already a result on
the ball park of the lattice one \cite{Csaki:1999vb} whilst the
sub-leading terms in the WKB approximation
(the rest of terms in r.h.s. of \bref{massqcd3}) is responsible for the
tiny difference between them.

The relation \bref{models} is not satisfied by the infinite square
well approximation with
$V_{well}=0$ \bref{sqwell}, indicating that the absolute normalisation
of the potential and
its width might play a crucial physical role. If this were not the case, our
regularised MN model would not only differ in the UV region, but
as a novelty
also in the deep IR.

\begin{figure}[t]
\begin{center}
\includegraphics[width=10cm,angle=-90]{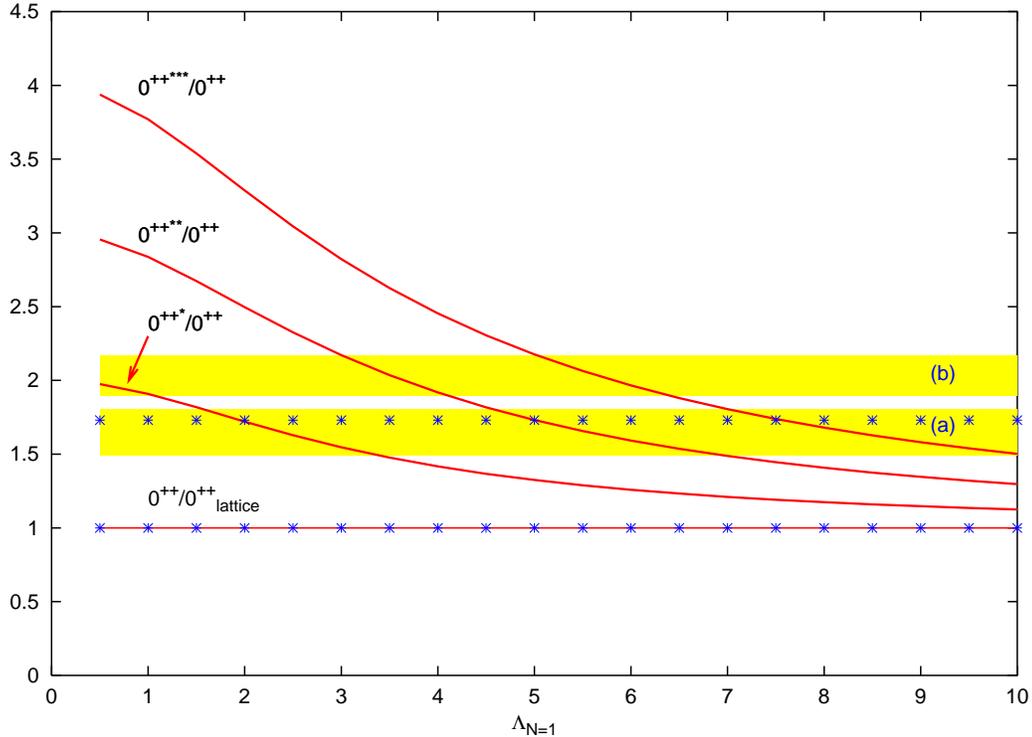}
\end{center}
\caption{\label{figlueballs}
Predictions of glueball masses as a function of
 $\Lambda_{{\cal N}=1} $ (solid lines).
Bands (a) and (b) correspond to edge values of
$M_{0^{++*}}/M_{0^{++}}$ and $M_{0^{++**}}/M_{0^{++}}$, respectively,
for the models quoted in table \ref{comp}.  Dots correspond to the ground
and first state lattice data.
The ground state has been normalised to $1$ in all
the cases. The numerical uncertainties are estimated to be of the
order of $2\%$.
}
\end{figure}

\avall

In what follows we shall investigate further the consistency of
these findings.
And in order to estimate roughly the range of
applicability of the model we shall compare the obtained glueball masses as a
function of the $\Lambda_{{\cal N}=1} $ cut-off with the lattice predictions.

\subsection{Numerology}

Our numerical results have been obtained by discretising the
Schr\"odinger equation \bref{sch2} for the \bref{MNpot-original}
potential with the \bref{noubc} BC and solving for the eigenvalues
using standard techniques. We collect in fig. \ref{figlueballs}
the glueball spectra (in arbitrary units) as a function of the
cut-off, $\Lambda_{{\cal N}=1}$, for the first excited states.
{F}or comparison we have also depicted in the horizontal bands the
range covered between the different results for the rest of the
models while dots are the lattice data. As is clear, from the
figure, once we push the cut-off to higher and higher-energy all
the glueball masses become degenerate and tend to vanish, $M_n
\sim n/\Lambda_{{\cal N}=1}$, thus recovering our earlier
expectations of the absence of glueballs in the original model 
\footnote{This effect also occurs in the KS model, once one pushes
the energy and cross to the UV region the spectra becomes continuous. We
thank R. Hen\'andez to point out this fact to us.}.
These considerations are complementary to the arguments cited
in the preceding subsection regarding the absence of discrete
spectrum for solutions of the dilaton equation \bref{dil-MN}
covering the full range of values from $\rho=0$ to $\rho
\rightarrow \infty$. Notice that the sensitivity of the
glueball masses with respect to the cut-off is rather large: 
to
use a modified cut-off accounting for the change from $\rho\approx 0.5$ to
$\rho\approx 5$ 
drives the glueball mass of the first two excited states
down by roughly $50 \%$.

At this point,
we are concerned that there is some degree of
arbitrariness on how to fix the value of the upper bound for
$\rho$, $\Lambda_{{\cal N}=1}$.
Nevertheless we have just seen that for a value approximately
of $\rho \approx 5$ we can safely consider that the R-symmetry has been
restored and hence the theory has changed the phase. An eyeball estimate from
the first excited state of
fig.~\ref{figlueballs} sets, comparing to the lattice results and 
restoring units,
$\Lambda_{{\cal N}=1}\approx 2\, l_s$,
as the scale of the effective theory that
the MN model describes.
Setting this scale one obtains the following spectra
normalised to $1$
\begin{equation}
M_1 = 1\,,\quad M_2 \approx 1.7\,,\quad
M_3 \approx 2.5\,, \quad M_4 \approx 3.3\,,\ldots\,,
\end{equation}
to be compared with table~\ref{comp} and with the crude estimate \bref{crude}
using $V_{\rm well}\simeq 1$, \bref{sqwell},
\begin{equation}
M_1 = 1\,,\quad M_2\approx 1.8\,,\quad
M_3\approx 2.6\,, \quad M_4\approx 3.4\,,\ldots\,.
\end{equation}
{F}rom the same figure, obtained with the exact potential,
the approximated pattern of the rest of supergravity
models can be recovered for the first few states
taking $\Lambda_{{\cal N}=1}\approx 3.5 l_s$,
\begin{equation}
M_1 = 1\,,\quad M_2\approx 1.5\,,\quad
M_3\approx 2.0\,, \quad M_4\approx 2.6\,,\ldots\,.
\end{equation}

As in the QCD$_3$ case we can apply the WKB approximation to
control the numerics. Although the approximation is valid for
excited states, there is already good agreement with the exact
results even for the first lowest ones differing with those by
less than $5\%$.

As a last point we  comment on the inclusion of non-commutative
effects \cite{nc}. These are identical to the ones described in
sec. \ref{n=0}: non-commutative effects increase the masses of the 
bound states. However, in this precise case the total number of
glueballs remains unlimited due to the presence of the infinite walls.

\avall\avall

We have interpreted the lack of glueball masses in the MN as an
explicit way of obtaining \emph{by hand} an estimate of the range of its
applicability. We shall explore in the next two sections further implications
of this frame.

\subsection{$1^{--}$ states}

We consider in what follows the linear
fluctuations of the RR 2-form potential $C_2$. We work in the
string frame and use the notation of \cite{joe} for the type--IIB
supergravity. It turns out that, due to the fact that in the MN
background the potentials $B_2$, $C_0$ and $C_4$ vanish, we can
get a solution for the linearised fluctuations in this background
where only the fluctuation $\tilde C$ of the $C_2$ potential is
turned on. The equation of motion  is
\begin{equation}
\partial_\mu\left(\sqrt{g} g^{\mu \alpha}g^{\nu \beta}
g^{\rho \gamma}\partial_{[\alpha}{\tilde C}_{\beta\gamma ]}\right)
=0 \,.
\end{equation}
Assuming the simplest ansatz we take only one component (in the
space directions) of the fluctuation, say $\tilde C_{12}$, to be
different from zero. Furthermore we consider the dependencies
\begin{equation}
\tilde C_{12} = f_{12}(\rho)e^{ik\cdot x} \label{ansatz-c}
\end{equation}
where $x^\mu$ runs over the 4-dimensional directions, $\mu =
0,\dots,3\,,$ which we take Euclidean. Then we obtain
\begin{eqnarray}
&&\frac{1}{Ng_s}\partial_\rho(e^{2\Phi_D(\rho)}e^{2g(\rho)}
\partial_\rho\tilde C_{12}) +\partial_\mu(e^{2\Phi_D(\rho)}e^{2g(\rho)}
\partial_\mu\tilde C_{12}) \nonumber \\ &&-
\partial_1(e^{2\Phi_D(\rho)}e^{2g(\rho)}
\partial_1\tilde C_{12}) -\partial_2(e^{2\Phi_D(\rho)}e^{2g(\rho)}
\partial_2\tilde C_{12}) = 0 \,,
\end{eqnarray}
together with
\begin{equation}
\partial_1\partial_0\tilde C_{12}
=\partial_1\partial_3\tilde C_{12}=\partial_1\partial_\rho\tilde
C_{12}= \partial_2\partial_0\tilde
C_{12}=\partial_2\partial_3\tilde C_{12}
=\partial_2\partial_\rho\tilde C_{12} = 0 \,.
\end{equation}
Requiring, according to the central idea in the ansatz
\bref{ansatz-c}, that $\tilde C_{12}$ has a sensible $\rho$
dependence, we are led to the result $k_1=k_2=0$. Finally, we end
up with
\begin{equation}
\label{1--}
\partial_\rho \left( e^{2\Phi_D(\rho)}e^{2 g(\rho)}
\partial_\rho f^{\mu \nu}(\rho) \right)
+ M^2 e^{2\Phi_D(\rho)}e^{2 g(\rho)} f^{\mu \nu}(\rho) = 0 \,,
\end{equation}
where $M^2 = - (k_0)^2 - (k_3)^2$ (with Euclidean time). This
result exactly coincides with the analogous equation for the
singlet dilaton fluctuations. Therefore the same conclusions
obtained there hold in this case; in particular, there is no 
discrete spectrum of normalisable states if we look for solutions
covering the
full range of the transversal variable. The same way to fix this
problem as in the case of the dilaton can be used here, and so we
end up with a degenerate spectrum for the dilaton fluctuations and
the RR 2-form fluctuations.

\subsection{Kaluza--Klein modes of $S^2$}
\label{kk}

The scenario presented in \cite{mn} possesses
many of the desired features one would like to find in QCD-like
theories as chiral symmetry breaking and the correct $\beta$-function
that matches precisely the one obtained from \CN=1 SYM field theory.
In addition,
in order to deal with a well-defined low-energy effective theory,
it is also desirable that the theory contains a mass gap between the
singlet (physical) and non-singlet states, being the latter heavier
enough to decouple. By themselves, their spectrum can be either
continuum or discrete.
It is also well known that qualitative arguments indicate some
contamination from spurious non-singlet
states obtained after the reduction of the $S^2$
that should properly have decoupled in the IR
$$
M^2_{\rm {glueballs}} \sim M^2_{\rm {KK}} \sim \frac{1}{N g_s
\alpha^\prime}\,,
$$
and as a consequence they do not
decouple in the limit of the validity of the supergravity
description, that requires $N g_s \gg 1$. Nevertheless, there is
evidence \cite{Pons:2003ci} that going beyond the supergravity
approximation (in the sense of introducing string solitons), the
analogue stringy KK modes on the $S^2$ sphere indeed decouple for
large values of the R-charge. We shall in the remainder analyse
the fate of the supergravity (point-like) KK modes in more detail
and check whether the decoupling observed for the stringy KK modes
still holds.

We use the complete ansatz \bref{ant} for the fluctuations of the
dilaton with modes on the $S^2$. As a matter of fact, one does not
need to consider modes on the $S^3$ because the 3-cycle appears
when we uplift our solution of 7-dimensional supergravity to 10
dimensions. Taking into account the truncation from type--IIB to
the 7-dimensional supergravity, this means that we can have
solutions for the fluctuations starting from the 7-dimensional
solution and then uplifting it: no fluctuation modes will be
turned on the $S^3$ if we do not perform further transformations
on the solutions.

The inclusion in the present ansatz of dependencies on the
variables of the $S^2$ makes it necessary to compute the
components $g^{\theta\theta}$ and $g^{\varphi\varphi}$ of the
metric. It turns out that these components are not affected by the
presence of the gauge fields $A^a$ in \bref{metric}.
The equation of motion for the non-singlet states reads
\begin{equation}\label{dil-MN-KK}
\partial_\rho(e^{2\Phi_D(\rho)}e^{2g(\rho)}\partial_\rho
\tilde{\Phi})+ \left(e^{2\Phi_D(\rho)}e^{2g(\rho)} N \alpha' g_s
M^2 - l(l+1) e^{2\Phi_D(\rho)} \right) \tilde{\Phi}=0\,,
\end{equation}
and one can already observe that the sign of the angular momentum contribution
is opposite to the $M^2$ one,
in a similar way as in section \bref{kk-qcd3}.
The potential obtained from \bref{dil-MN-KK} has a centrifugal barrier
similar as in 3-dimensional central potentials,
$$
V(\rho,M,l) = {\rm eq}.~\bref{MNpot-original} +
8 l (l+1) \frac{\sinh^2(2\rho)}{1-8\rho^2-\cosh(4\rho)+4\rho\sinh(4\rho)}\,.
$$
The term with the angular momentum dependence behaves as
$\sim 1/\rho^2$ for
$\rho \rightarrow 0$ and tends to zero for large $\rho$. 
It is again obvious that the shape of this potential, steadily
falling from infinity at $\rho =0$ to the asymptotic value $1-M^2$
at $\rho \rightarrow \infty$, will prevent the model from
exhibiting a discrete spectrum of normalisable wave functions. 
We can still adopt the
same drastic procedure of imposing a hard cut-off, as we did for
the glueballs. Since the potential is sensitive to the singularity only
very close to the origin, one expects
that the mass spectrum for the first
excited states being of the same order of magnitude as for the glueballs
spectrum. We have verified this claim numerically.
Indeed, the KK masses for $l=1$,
using $\Lambda_{{\cal N}=1}=2$ and normalising
to the singlet $0^{++}$ ground state, are
\begin{equation}
M_1 \approx 2.4\,,\quad M_2\approx 3.7\,,\quad
M_3\approx 5.1\,, \quad M_4\approx 6.5\,,\ldots\,.
\end{equation}
We conclude that, within this interpretation of the
model, which allows for the introduction of the hard cut-off at
the scale of the R-symmetry breaking, there is no decoupling of
the KK supergravity modes.

\section{Summary and discussion}
\label{summary}
\setcounter{equation}{0}

Extending the AdS/CFT correspondence to non-supersymmetric
and/or non-maximally supersymmetric theories we have computed the
glueball spectrum for the non-commutative QCD$_3$ and for \CN=1 SYM
Maldacena--N\'u\~nez model. We have assumed that the same relation
as in AdS/CFT holds in these cases, and thus we have identified
the glueball spectrum with the one coming from the dilaton in
type--IIB strings.

In the case of QCD$_3$ the inclusion of a non-commutative
parameter, $\theta$, does not improve the lack of decoupling of
the non-singlets KK states with respect to the singlet ones. As a
general feature of this model we see that the
potential reaches its minimum at $\theta=0$. {F}or $\theta \ne 0$
the parameter plays the same role as the strength of the electric
potential in a cool emission of electrons from a metal: in
increasing $\theta$ the number of bound states decreases. It
exists a critical value of $\theta$ where there is no longer any
bound state in the system, see fig.~\ref{fig1}.

The most relevant findings are concerning the glueball spectra of
the MN model. To obtain the spectra we looked at the poles of
massless dilaton two-point function: in order to get any n-point
Green function one shall simply take n-functional derivatives on
the generating functional defined in \bref{corre} considering the
fields $\phi$ as external sources for the operators ${\cal O}$.
{F}or the two-point Green-function in momentum-space case this
gives
\begin{equation}
\big{\langle} {\cal O}_4(l) {\cal O}_4(m) \big{\rangle} =
\frac{\partial^2 S[\phi]}{\partial\phi_l \partial\phi_m}\,.
\end{equation}
Taking the minimal massless action for the fluctuations of the
scalar field in the IR region of the MN background (in the string frame)
\begin{equation}
S=\frac{1}{4\kappa_{10}^2}\int d^{10}\,x\,
e^{3\Phi_D(\rho)}e^{2g(\rho)} g^{\mu\nu}
\partial_\mu\phi
\partial_\nu\phi
\end{equation}
with all the rest of fields set to zero and integrating by parts,
one can identify easily the dilaton e.o.m.
\bref{dil-MN} with $M=0$. If we further set to zero all
the boundary terms coming from the $\mathbb{R}^4$ integration the
action takes the form
\begin{equation}
S = \frac{1}{16\kappa_{10}^2} \left(\frac{N}{4}\right)^{3/2}
\int_{S^3} d\Omega_3^2 \int_{S^2} d\Omega_2^2 \int_0^{\Lambda_{{\cal
N}=1}}
d^4\,x \left( {\cal F}(x^\mu,\rho)-{\cal F}(x^\mu,0) \right)\,,
\end{equation}
where we have defined the flux factor ${\cal F}(x^\mu,\rho) \equiv
e^{3\Phi_D(\rho)}e^{2g(\rho)}\phi(x^\mu,\rho) \,\partial_\rho
\phi(x^\mu,\rho)$. Notice that even at this earlier stage 
one can realise that the previous flux factor can not be extended
to the full range of $\rho$ if its value has to be bounded, signaling
that we have to change to the S-dual version of the metric.

In order to find a discrete spectrum for normalisable glueball states 
we have first considered the problem of the runaway of the dilaton. 
Since this causes the supergravity approximation to lose its validity, 
we are led to move at some moment to the S-dual solution. 
This means that we shall use \bref{metric} in the IR region together
with the BC $\Psi(0)=0$ and \bref{smetric}  with the BC
$\Psi(\rho\rightarrow\infty)= 0$ at the UV. Notice that there is
an intermediate region where both descriptions fail to
give a reliable description, shadow area in fig.~\ref{phs}.
The equation for the fluctuations for the dilaton remains the same in both 
regions and the only change is in the definition of the norm of our wave 
function. 
It turns out that all solutions become normalisable and consequently there 
is no discrete spectrum. This
conclusion is in flat disagreement with the findings through the
Wilson loop \cite{mn}. To disentangle this puzzle we noticed that:
{\it i)} the expectation that the existence of an area law in the
Wilson loop, as defined by \cite{rey,wilson}, does not necessary
implies the existence of the glueball spectra in our case. This is
probably not only due to the non-conformal nature of the model but
as a consequence of the dilaton blowing up at the UV region. {\it
ii)} One can amend the model by setting \emph{by hand} a
hard cut-off identified with the R-symmetry scale, $\Lambda_{{\cal
N}=1}$. Above the cut-off the model makes no sense whilst below
can be interpreted as an effective description of \CN=1 SYM. In
doing so we enforce the existence of the spectra with a
distribution rather similar to the one of a square-well potential
in quantum mechanics. This can be justified to a considerable
extent, on the basis of the R-symmetry breaking of the underlying
field theory. Within this regularisation and relying on the
lattice values of the two lightest $0^{++}$ states, we have set
the scale of  applicability of the model,  $\Lambda_{{\cal
N}=1}\approx 2\,l_s$. The lack of lattice results for higher
excitations prevents us from checking the consistency of these
results. As a consequence of this choice, notice that the MN model
does not only differ from similar models (KS, \CN=1$^*$) in the UV
region as mentioned earlier in the literature but also in the IR.
We should point out that, while the glueball spectra in the rest of
supergravity models are in reasonable agreement among themselves,
the MN one clearly does not follow this general trend. Otherwise,
we can recover a pattern similar to the rest of supergravity
models discarding the previous value of $\Lambda_{{\cal N}=1}$. In
this case a value $\Lambda_{{\cal N}=1}\approx 3.5 \,l_s,$ is
favoured.

Using the same approach, and to test further its implications, we
have computed the $1^{--}$ states spectra. It turns out that it
reduces to the RR 2-form $C_2$ and is degenerate with $0^{++}$.

We have also studied the KK modes appearing on the $S^2$. Like in
the previous two cases there is need for regularisation.
After this, it is unlikely that the model presents
decoupling.

All in all, we point out that the MN model does not display a 
discrete glueball
spectra by itself. With a rude regularisation we interpret it as a
low-energy effective theory and estimate its range of validity,
that turns out to be quite limited. Even though we do not claim
that the amended model is ruled out, it is hard to believe that it
gives any sensible results, unless one finds a precise
determination, coming from the model alone --or some improvement
of it--, of the value of the cut-off $\Lambda_{{\cal N}=1}$.

\vskip 6mm
{\it{\bf Acknowledgements}}

We are grateful to Rafael Hern\'andez and Carlos N\'u\~nez
for useful discussions and comments.
This work is partially supported by MCYT FPA, 2001-3598, CIRIT, GC
2001SGR-00065, and HPRN-CT-2000-00131. J.M.P. acknowledges the
Spanish ministry of education for a grant.

\vskip 4mm

\appendix
\renewcommand{\theequation}{\Alph{section}.\arabic{equation}}
\section{Schwarzian derivative and reparameterisation}
\label{app} \setcounter{equation}{0}

Aiming at the determination of the spectrum of
masses $M$, it is convenient to write the equation as a
zero-energy one-dimensional quantum mechanical problem,
\begin{equation}
\label{zeroenergy} \frac{d^2 \psi}{d u^2} - V(u) \psi = 0\,,
\end{equation}
where the potential $V(u)$ already carries the parameter $M$ whose
spectrum we are looking for.

To bring a general equation of the type,
\begin{equation}
\label{genequ} \partial_u(f(u)\partial_u \phi) + g(u)\phi =0\,,
\end{equation}
to the form \bref{zeroenergy} we use a multiplicative factor for
$\phi$ to define a new function $\psi := (f(u))^{\frac{1}{2}}
\phi$. The
 potential obtained has already been written in \bref{pot1},
 with $g(u) = M^2 h(u) + p(u)$.

Notice that we still have to our avail the freedom of
reparameterisation, that is, of changing
the coordinate system, $u \rightarrow \omega, \, u=F(\omega)$.
This change will bring an equation of the type \bref{zeroenergy}
again to the general form \bref{genequ}, but a new multiplicative
factor will transform the equation into the Schr\"odinger form,
with a new potential $\tilde{V}(\omega)$. A simple computation
shows that the new potential is related to the old one by
\begin{equation}
\label{newpotential} \tilde{V}(\omega) =
V(F(\omega))(\partial_\omega F)^2 - \frac{1}{2} Q_F(\omega)\,,
\end{equation}
where \begin{equation} \label{schwartzian}
  Q_F(\omega) :=
\left(\frac{F''}{F'}\right)'-\frac{1}{2}\left(\frac{F''}{F'}\right)^2
\end{equation}
is the Schwarzian derivative of  $F$ ($F'$ stands for $\partial_w
F$).

The reason for the appearance of the Schwarzian derivative
becomes clear when one takes into account that composition of two
reparameterisations is a closed operation resulting in a third one
and that, under such a composition, the Schwarzian derivative
satisfies the key property
\begin{equation}
  Q_{F\circ G}(\omega) =  Q_F(G(\omega))\, (G'(\omega))^2+ Q_G(\omega) \,.
\end{equation}
Indeed, under an infinitesimal coordinate transformation
$u\rightarrow u - \epsilon(u)$, the potential transforms as
\begin{equation} \label{infinit-transf}
  \delta V =  \epsilon V' + 2 \epsilon' V - \frac{1}{2}\epsilon'''\,,
\end{equation}
which is the standard transformation rule for a weight 2 scalar
density under reparame\-te\-ri\-sations  with an additional
``anomalous'' term $\epsilon'''$, which is compatible with the
algebra of reparame\-te\-ri\-sations. Notice that the coefficient in
front of this term can be arbitrarily varied by just redefining
$V$ in \bref{infinit-transf} with a numerical factor. With
$\frac{c}{12}$ instead of $-\frac{1}{2}$, \bref{infinit-transf}
matches the transformation rule for the energy-momentum
tensor of two-dimensional quantum conformal field theories, where
$c$ is the conformal anomaly.

The result \bref{newpotential} provides us therefore with an
indirect method to integrate the infinitesimal transformation
\bref{infinit-transf}. A direct method, which essentially
exponentiates the function $\epsilon$, can be found in
\cite{Gracia:1993wn}.

\section{On the WKB method and reparameterisation}

Let us make a comment on the WKB method, when applied to
the determination of the energy spectrum of bound states. Suppose
we are given a Schr\"odinger equation with a potential ${\cal V}$
suitable for the application of the method, with two turning
points, $x_1,\, x_2$, for the equation $E-{\cal V}(x) =0$ in a
certain interval of values of the energy $E$. The WKB formula for
the spectrum of energies is
\begin{equation}
\label{wkbsencera} I(n)\equiv \left( n +\frac{1}{2}\right) \pi =
\int_{x_1}^{x_2} dx \sqrt{E -{\cal V}(x)}\,, \quad n=0,\, 1,\,2,\,
\dots \,.
\end{equation}
Now let us introduce an infinitesimal reparameterisation
\footnote{Reparameterisations in the WKB method
were first considered in \cite{miller}.}, $x
\rightarrow x + \epsilon(x)$. One can compute, to first order in
$\epsilon$, the new potential for the new Schr\"odinger-like
equation. Since the transformation is infinitesimal, it is still
an eigenvalue problem, the new potential being ${\cal \tilde V} =
{\cal  V} + \delta {\cal  V}$, with $\delta {\cal V}$ as in
\bref{infinit-transf}. Accordingly, the turning points undergo a
change,
\begin{equation}
x_{0,1} \rightarrow {\tilde x}_{i} = x_{i} -\epsilon(x_{i}) +
\frac{1}{2} \frac{\epsilon'''(x_{i})}{{\cal  V}'} \,,
\end{equation}
whereas the new expression for \bref{wkbsencera} becomes
\begin{eqnarray}
\label{wkbepsilon} I(n)\equiv \left( n +\frac{1}{2}\right) \pi
 &=& \int_{{\tilde x}_1}^{{\tilde x}_2} dx \sqrt{E
-{\cal \tilde V}(x)} \nonumber \\ &=& \int_{x_1}^{x_2} dx \sqrt{E
-{\cal V}(x)} + \frac{1}{4}\int_{x_1}^{x_2} dx
\frac{\epsilon'''(x)}{\sqrt{E -{\cal V}(x)}}\,, \quad n=0,\,
1,\,2,\, \dots \,. \nonumber\\
\end{eqnarray}
We have the previous expression but with a new contribution to
the r.h.s. Here we see in this last term an unwanted correction to
the WKB formula, originated from the Schwarzian derivative term.
Should the WKB formula be an exact one no matter the coordinate
system, this last term would not exist. But this correction is
unavoidable for general reparameterisations and in a certain sense
reflects the fact that the WKB formula is only approximate. For
instance, in the particular case where it gives the exact result
for the spectrum when using, say, the $x$ coordinates, the
spectrum will no longer be the exact one when using the new
coordinates ${\tilde x}$. At first sight this might seem strange,
even dangerous, as a paradox built in the WKB method, but it is
just an outcome of its approximate nature. Notice, however, the
important point one can deduce from \bref{wkbepsilon}: for
energies $E$ sufficiently large, this correction term becomes
negligible. This valuable result is in the line of the usual
assertion that for large quantum numbers the errors given by the
WKB method are small. We think that this argument can be extended,
by iterations of the infinitesimal transformation, to general
reparameterisations that yield potentials still suitable for the
application of the WKB method, and that the correction term to the
WKB formula due to the Schwarzian derivative will always be
sub-leading with respect to the main term.

\section{Normalisation for the Sch\"odinger wave function}

In our procedure to deal with equation \bref{sch} through WKB methods, we
have reformulated and converted it into a Schr\"odinger-like
equation. It is therefore necessary to discuss the normalisation
of the corresponding wave function. To this end we must first
write down the normalisation associated with \bref{laplace}. The
requirement of hermiticity of the Laplace operator induces the
natural normalisation, in the Einstein frame,
\begin{equation}
 \int d\rho \sqrt{g}\vert\Phi\vert^2 \,.
\end{equation}
Now consider the steps taken to bring equation\bref{eq1bis} to a
Schr\"odinger-like form, with  $ f =\sqrt{g} g^{\rho\rho}$ and
$\Psi = f^{1/2}\Phi$ ,
\begin{equation}
 \int d\rho\,\sqrt{g}\vert\Phi\vert^2 =
 \int d\rho\,\sqrt{g}f^{-1}\vert\Psi\vert^2
= \int d\rho\, g_{\rho\rho}\vert\Psi\vert^2\,. \label{norma}
\end{equation}
This last expression is the normalisation we were seeking for.
Under the reparameterisation $\rho = F(\omega)$, and with
$\tilde{A}$ representing the transformed of a generic object $A$,
the only component $g_{\rho\rho}$ of the one-dimensional metric
$g_{\rho\rho} d\rho^2$ behaves as a weight=2 scalar density,
$\tilde{g}_{\omega\omega} = g_{\rho\rho}(F')^2$, whereas the wave
function $\Psi$ behaves as a weight=(-1/2) scalar density,
$\tilde{\Psi} =  \Psi (F')^{-1/2}$, thus producing the invariance
\begin{equation}
 \int d\rho\, g_{\rho\rho}\vert\Psi\vert^2 =
 \int d\omega\, \tilde{g}_{\omega\omega}\vert\tilde{\Psi}\vert^2\,.
\end{equation}
We conclude that our one-dimensional Schr\"odinger problem
includes two pieces of information, the Schr\"odinger-like
equation itself, \bref{zeroenergy}, and a one-dimensional metric,
$g_{\rho\rho} d\rho^2$, originated in \bref{laplace} that plays a
role in the definition of the wave function normalisation.

\end{document}